\documentclass[aps,twocolumn,groupedaddress,showpacs]{revtex4}
\usepackage{amsmath,amscd,amssymb,epsfig}
\begin{document}
\bibliographystyle{apsrev}



\title[]{Viscoelastic subdiffusion: from anomalous to 
normal}

\author{Igor Goychuk}

\affiliation{
 Institut f\"ur Physik, Universit\"at Augsburg,
  Universit\"atsstr. 1,
  D-86135 Augsburg, Germany
}

\date{\today}

\begin{abstract}
 
We study viscoelastic subdiffusion in bistable and periodic 
potentials within the Generalized Langevin Equation approach. Our
results justify the  (ultra)slow fluctuating rate view of the
corresponding bistable non-Markovian dynamics which displays bursting
and anti-correlation of the residence times in two potential wells. 
The transition kinetics is asymptotically stretched-exponential when
the  potential barrier $V_0$ several times exceeds thermal energy
$k_BT$  ($V_0\sim 2\div 10\; k_BT$) and it cannot be described  by the
non-Markovian rate theory (NMRT).  The well-known NMRT result 
approximates, however, ever better with the increasing barrier height,
the most probable logarithm of the residence times.   Moreover, the
rate description is gradually restored when the barrier height exceeds
a fuzzy borderline which depends on the power law exponent 
of free subdiffusion $\alpha$.  
Such a potential-free subdiffusion is ergodic. 
Surprisingly, in  periodic potentials it is not sensitive to the 
barrier height in the long time asymptotic limit.  However, the
transient to this asymptotic regime is extremally slow and it does
profoundly depend on the barrier height.  The time-scale of such
subdiffusion can exceed the mean residence time in a potential well,
or in a finite spatial domain by many orders of magnitude. All these
features are in sharp contrast with an alternative subdiffusion
mechanism involving jumps among traps with the divergent mean
residence time in these traps.

\end{abstract}


\maketitle



\section{Introduction} Multifaceted anomalous diffusion attracts ever
increasing attention, especially in the context of biological
applications. For example, diffusion of mRNAs and ribosomes in the
cytoplasm  of living cells is anomalously slow \cite{Golding}, large
proteins behave  similarly \cite{Saxton,Tolic,Weiss1,Banks}.  Even
intrinsic conformational dynamics of the  protein macromolecules can
be subdiffusive \cite{McCammon,Bizzarri,Kneller,Luo,Neusius,
Yang1,Granek,Min1,Min2,Goychuk04}.  There is a bunch of different
physical mechanisms and the corresponding theories attempting to
explain the observed  behaviors, from spatial and/or time fractals, 
influence of disorder, cluster percolation, etc.,  to viscoelasticity
of complex media
\cite{Feder,Hughes,Scher,Shlesinger,Bouchaud,Havlin,Dewey,Metzler,
Amblard,Mason,Qian,Condamin,Caspi}. In particular, molecular crowding
can be responsible for the viscoelasticity of dense suspensions like
cytosol of bacterial cells lacking a static cytoskeleton
\cite{Banks,Mason,Guidas}.  The state of the art remains rather
perplexed, offering cardinally different  views on the underlying
physical mechanisms, as we clarify further with this work. 

One physical picture  reflects a  set of the traps (possibly dynamical
\cite{Condamin}) where the  diffusing particle stays for a random time
$\tau_i$.  The mean residence time (MRT) in traps should diverge
\cite{Scher,Shlesinger}  for the  diffusion to become anomalously slow, i.e.
with the position variance growing sublinearly,  $\langle \delta
x^2(t)\rangle\sim t^{\alpha}$, with $0<\alpha<1$.  This stochastic
time-fractal picture  became one of the paradigms in the field
\cite{Hughes}. It can be also related to averaging over static, or 
quenched disorder \cite{KlafterSilbey,Hughes}. Such a continuous time
random walk (CTRW) among traps  has infinite memory even if the
residence times in the different traps are not correlated.  The memory
comes from the non-exponential residence time distributions in traps. 

A quite different physical view was introduced  by Mandelbrot {\it et
al.} \cite{Mandelbrot,Feder} with  the fractional Brownian motion
(fBm).  Here, the standard Gaussian Wiener process with independent
increments is generalized to  incorporate the statistical dependence
of increments.  The Gaussian nature remains untouched, but the
increments can be either positively, or negatively correlated over an
infinite range.  Positive correlations (persistence) lead to
superdiffusion.  If correlations are anti-persistent, i.e. given a
positive increment, the next one will, with greater probability, be a
negative increment and vice versa, then a subdiffusive behavior can
result. This idea does not imply that the  residence time in a finite
spatial domain diverges on average.  Here roots the {\it cardinal}
difference, in spite of some superficial similarities, between the
fBm-based  and the CTRW-based approaches to subdiffusion. 

FBm emerges naturally, e.g.,  in viscoelastic media as one of the best
justified models. Indeed, let us start from a phenomenological
description of viscoelastic forces acting on a particle moving with
velocity $\dot x(t)$ in some time-window $[0,t)$:
\begin{equation}
F_{v-el}(t)=-\int_{0}^{t}\eta(t-t')\dot x(t')dt'.
\end{equation}
Clearly, for a memory-less {\it linear} frictional kernel with
$\eta(t)=2\eta\delta(t)$ on the particle  acts a purely viscous Stokes
friction force, $F_v=-\eta\dot x$. If memory does not decay,
$\eta(t)=\eta=const$, then the  force  is quasi-elastic,
$F_{el}(t)=-\eta[x(t)-x(0)]$ (cage force). A popular model of
viscoelasticity introduced by Gemant \cite{Gemant},  which
interpolates between these two extremes, corresponds to 
$\eta(t)=\eta_{\alpha}t^{-\alpha}/\Gamma(1-\alpha)$ with $0<\alpha<1$
($\Gamma(x)$ is the gamma-function).  Remarkably, this  model yields
the Cole-Cole dielectric response for particles trapped in parabolic
potentials \cite{Cole,GoychukRapid}, which is frequently observed in
complex media. For a small Brownian particle of mass $m$, one must
take into account unbiased random forces $\xi(t)$ acting from the 
environment (Langevin approach).  Then, the {\it linear} friction
approximation  combined with the symmetry of detailed balance
fixes the statistics of the stationary {\it thermal} random forces to
be Gaussian  \cite{Reimann}. Moreover,  the fluctuation-dissipation
theorem dictates   that the stationary autocorrelation function of the
random force,  temperature $T$, and the memory kernel are related
\cite{Kubo}:
\begin{equation}\label{FDT}
\langle \xi(t)\xi(t')\rangle=k_BT\eta(|t-t'|) \;.
\end{equation}
For Gemant model, $\xi(t)$ is the 
fractional Gaussian noise (fGn) \cite{Mandelbrot,Min1,GoychukRapid}. 
Altogether, the motion in potential $V(x,t)$ 
is described by the generalized 
Langevin equation (GLE)
\begin{equation}\label{GLE}
m\ddot x+\int_{0}^{t}\eta(t-t')\dot x(t')dt'+\frac{\partial V(x,t)}{\partial x}
=\xi(t) \;.
\end{equation}
Importantly, this GLE can also be derived from the mechanical 
equations of motion for a particle
interacting with a thermal bath of harmonic oscillators, i.e. from 
first principles. 
 This  statistical-mechanical derivation 
 \cite{Bogolyubov,Kubo,Zwanzig,WeissBook} 
 involves the spectral 
 density $J(\omega)$ of bath oscillators. It is related to the  
spectral density of thermal force, 
$S(\omega)=2\int_{0}^{\infty}\langle \xi(t)\xi(0)\rangle\cos(\omega t)dt$,
as $S(\omega)=2k_BTJ(\omega)/\omega$. 
With $J(\omega)=\eta_{\alpha}\omega^{\alpha}$, the
so-called Ohmic case of $\alpha=1$ corresponds to viscous Stokes friction and normal diffusion.
The sub-Ohmic, or fracton thermal bath  \cite{WeissBook,Granek}
with $0<\alpha<1$ and $1/f^{1-\alpha}$ noise 
spectrum of random force corresponds to the above Gemant model
of viscoelasticity. It  
yields subdiffusion in the potential-free case \cite{Wang,WeissBook}.
The velocity autocorrelation function is then negative (except of origin), being
the reason for the anti-persistent motion. The physical origin of this feature is that  
the elastic component of the viscoelastic force
opposes the motion and ever tries to restore the current particle's position. 
Moreover, in the inertialess limit ($m\to 0$) the solution of GLE is 
fBm with the coordinate variance
$\langle \delta x^2(t)\rangle=2K_{\alpha}t^\alpha/\Gamma(1+\alpha)$ 
\cite{Caspi,PRL07} and the subdiffusion coefficient $K_{\alpha}$ obeying
the generalized Einstein relation
$K_{\alpha}=k_BT/\eta_{\alpha}$ \cite{Chen,MetzlerPRL}. 
This way, anti-persistent subdiffusive fBm 
emerges from first principles within a physically well grounded, but 
approximate description. 
It corresponds also exactly 
to the diffusion equation with a time-dependent diffusion coefficient 
$D(t)=K_{\alpha}/[\Gamma(1+\alpha)t^{1-\alpha}]$ \cite{Adelman} 
which is frequently used
to fit experiments in viscoelastic and crowded environments, see e.g. in 
\cite{Saxton,Banks,Guidas}.

\section{Theory}
\begin{figure}
\vspace{0.2cm}
\centerline{\includegraphics[width=7.5cm]{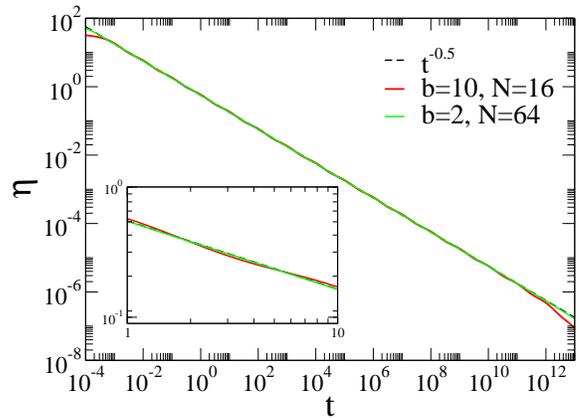}}
\caption{
(Color online) Frictional power law memory kernel (in units of $\eta_\alpha$) and its two different 
approximations  versus
time $t$ (arbitrary units) for $\alpha=0.5$ and $\nu_0=10^3$. Notice that the approximation
with $b=2$ practically coincides with the exact kernel in this plot and the choice $b=10$
is also  a very good one in spite of logarithmic oscillations which are barely seen.
Inset magnifies a part of plot.
}\label{Fig1}
\end{figure}

\begin{figure}[t]
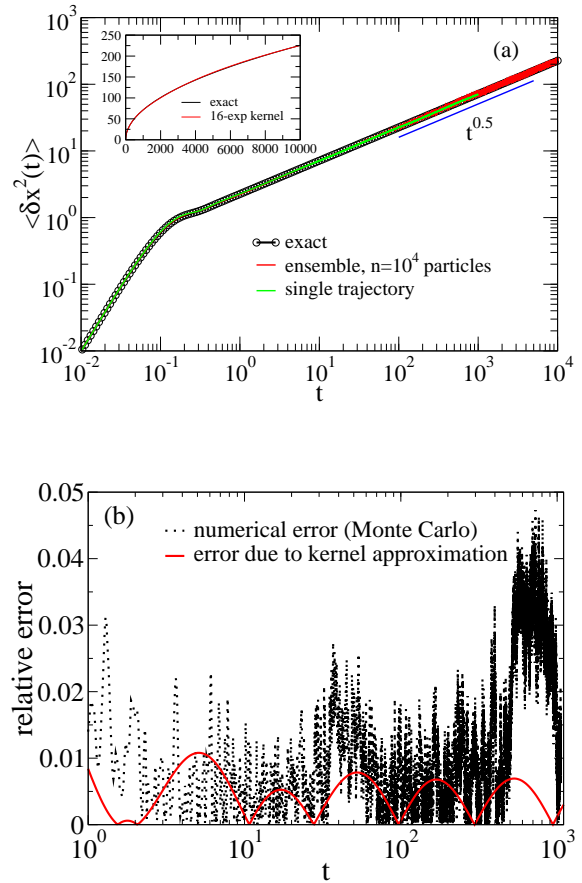

\vspace{0.5cm}
\centering
\includegraphics[width=7.5cm]{Fig2a.eps}

\vspace{1cm}

\includegraphics[width=7.5cm]{Fig2b.eps}
\caption{(Color online) (a) Stochastic simulations of Eqs. (\ref{embedding})-(\ref{eq6}) with embedding
dimension $D=18$ {\it vs.} exact solution of GLE (\ref{GLE}) for
free subdiffusion. Coordinate is scaled in some arbitrary units $L$ and time is scaled in units of 
$\tau_{\beta}=(L^2\eta_{\alpha}/k_BT)^{1/\alpha}$; $\alpha=0.5$ and the 
damping parameter $r_{\beta}=\tau_{\beta}/\tau_{\rm bal}=10$
with $\tau_{\rm bal}=L/v_T$ and $v_T=\sqrt{k_B T/m}$. 
Particles are initially localized at $x=0$ with Maxwellian
distributed velocities. The number of particles $n=10^4$ is used in simulations
(stochastic Heun algorithm, time step $\Delta t=10^{-4}$) for 
the ensemble averaging. The time averaging for a
single trajectory is done using Eq. (\ref{A5}). The agreement proves ergodicity. 
Inset in  compares two numerically exact 
solutions of GLE, one with the strict power law kernel and one with 
its $16$-exponential approximation. (b) Relative numerical error
of our simulations (versus the {\it exact} solution) is presented.
The relative deviation of the solution with the approximate memory 
kernel from the solution with exact power law kernel is also
presented. Notice the occurrence of logarithmic oscillations
within a less than 1\% error margin, which, however, do not
play any essential role. 
} 
\label{Fig2}
\end{figure}

Anomalous escape (rate) processes and spatial 
subdiffusion in periodic 
potentials represent within the GLE description a highly nontrivial, 
longstanding challenge. 
Even the corresponding non-Markovian 
Fokker-Planck description is generally not available, 
except for the strictly linear and parabolic potentials \cite{Adelman}.
To get insight into the physics of such processes, it is convenient 
to approximate $1/f^{1-\alpha}$ noise by a sum of independent  
Ornstein-Uhlenbeck
noise components, $\xi(t)=\sum_{i=0}^{N-1}\zeta_i(t)$, 
with the autocorrelation functions,
$\langle\zeta_i(t)\zeta_j(t')\rangle=k_BT\eta_i
\delta_{ij}\exp(-\nu_i|t-t'|)$. The corresponding memory kernel is accordingly 
approximated
by a sum of exponentials,
\begin{equation}\label{approx} 
\eta(t)=\sum_{i=0}^{N-1}\eta_i\exp(-\nu_i t),
\end{equation}
where $\nu_i=\nu_0/b^i$ is the inverse autocorrelation time  of the $i$-th
component and 
$\eta_i=(\eta_{\alpha}/\Gamma(1-\alpha))C_{\alpha}(b)
\nu_0^\alpha/b^{i\alpha}$ is its weight.
Furthermore, $\nu_0$ presents the high-frequency 
cutoff of $\xi(t)$, $b$ is a 
dilation (scaling) parameter, and $C_{\alpha}(b)$ is a numerical 
constant.
The low-frequency noise cutoff is $\omega_c=\nu_0/b^{N-1}$. 
It is worth to mention
that such cutoffs emerge for any $1/f$ noise on the physical 
grounds \cite{Weissman}. For $\alpha=0.5$, which is of experimental 
interest \cite{Min1,Min2}, the choice
of $b=2$ (i.e. octave scaling) 
and $N=64$ with 
$\nu_0=10^3$ (in arbitrary units) and $C_{1/2}(2)=0.389$ 
allows one to fit perfectly
the power law kernel 
in the range from $t=10^{-3}$ till $t=10^{15}$, i.e. over 18 time decades.
The choice of  $b=10$ (i.e. decade scaling) 
with $C_{1/2}(10)=1.3$ 
provides also an excellent fit over 15 time decades from 
$t=10^{-3}$ till $t=10^{12}$ with $N=16$.
 The numerical advantage of larger $b$ is that one can use 
smaller Markovian embedding dimension $D=N+2$. These two approximations to
the exact power law memory are shown in Fig. \ref{Fig1}. The approximation 
with $b=10$ displays logarithmic oscillations \cite{Hughes} which are 
barely seen in this plot and make a little influence on the stochastic
dynamics, see in Fig. \ref{Fig2}. 

Free subdiffusion holds
until the time scale of $1/\omega_c$ which can be very large. 
The idea of such a representation of a power law dependence 
is rather old \cite{Palmer,Hughes}, being also 
habitual in the $1/f$ noise theory \cite{Weissman}. 
The corresponding
power spectrum  
is approximated by a sum of Lorentzians, 
$S(\omega)=2k_BT\sum_{i}\frac{\eta_i\nu_i}{\omega^2+\nu_i^2}$. 
Every stationary noise component is asymptotic
($t\to\infty$) solution of 
$\dot\zeta_i(t)=
-\nu_i\zeta_i(t)+\sqrt{2\eta_i\nu_ik_BT}\xi_i(t)$, where $\xi_i(t)$ are
independent white Gaussian noises with unit intensity, 
$\langle \xi_i(t)\xi_j(t')\rangle=\delta_{ij}\delta(t-t')$. 
Furthermore, the particle must act back on
the source of noise in order to have the FDT relation (\ref{FDT}) satisfied. 
This yields the following  $D=N+2$ 
dimensional 
Markovian embedding of the non-Markovian
GLE stochastic dynamics in Eq. \eqref{GLE} with kernel (\ref{approx}): 
\begin{eqnarray}\label{embedding}
\dot x&=& v\;,\\
m\dot v & =& -\frac{\partial V(x,t)}{\partial x}+
\sum_{i=0}^{N-1}u_i(t)\label{eq5} \;,\\
\dot u_i& = &-\eta_i v-\nu_iu_i+\sqrt{2\nu_i\eta_ik_BT}\xi_i(t) \;.
\label{eq6}
 \end{eqnarray}
Initial $u_i(0)$ have to be sampled independently from unbiased Gaussian distributions 
with the standard deviations $\sigma_i=\sqrt{k_BT\eta_i}$ \cite{Kupferman}.
Under this condition, it is easy to show that  Eqs. (\ref{embedding})-(\ref{eq6})  are equivalent to
the GLE (\ref{GLE}) with kernel (\ref{approx}) under FDT relation (\ref{FDT}).
Notice that $u_i(t)$ are auxiliary mathematical variables and should not be interpreted
as (scaled) coordinates of some physical particles. The embedding (\ref{embedding})-(\ref{eq6})
can be also derived from a more general scheme in Ref. \cite{Kupferman}.

The numerical simulations of Markovian dynamics in Eqs. (\ref{embedding})-(\ref{eq6}) 
below were done with stochastic Euler and Heun algorithms 
\cite{Gard} using
different random number generators. 
The results are robust. The simulations have been checked against 
the exact analytical results available for the potential-free case and parabolic
potentials. The both above embeddings with $b=2, N=16$ and 
$b=10, N=64$ yield practically
the same results within statistical errors. 
However, the simulations with $N=16$ requires much less
computational time. 
Therefore, we preferred the latter $D=18$-dimensional 
embedding in most simulations  following a ``rule of thumb'' 
\cite{Sansom}: a negative power law 
extending over $n$ time-decades can be approximated 
by a sum of about $n$ exponentials. 
The quality of this approximation along with numerical
errors is discussed in the Appendix A and Fig. \ref{Fig2} by making comparison
of the numerical Monte Carlo results with the numerically 
exact solution of the free subdiffusion problem. The numerical error
is  mostly less than 3\%  for free subdiffusion in this work, 
whereas the theoretical error incurred
by the $16$-exponential approximation of the power law memory
kernel is mostly less than 1\%, cf. Fig. \ref{Fig2}. Clearly, it makes no sense to
approximate the kernel better, if no more than $n=10^4$ trajectories
are used in the ensemble averaging.

The chosen Markovian embedding of non-Markovian GLE dynamics 
with $\eta(t)$ in Eq. (\ref{approx})
is mathematically, of course, 
not unique. Another embedding was proposed in Ref. \cite{Marchesoni} and infinitely many different embeddings of one and the same non-Markovian dynamics 
are in fact possible  \cite{Kupferman}. 
However, our simplest scheme allows to contemplate straightforwardly
the view of anomalous escape processes as rate processes
with dynamical disorder  \cite{Min2,Zwanzig2,WangWolynes}. 

\subsection{Non-Markovian rate theory and beyond}
We consider now stochastic transitions  in a paradigmatic bistable quartic
potential $V(x)=V_0(1-x^2/x_0^2)^2$ with minima located at 
$x_{\rm min}=\pm x_0$ and the barrier height $V_0$. The question is: Which is
the statistical distribution of the residence times in two
potential wells and the escape kinetics?
This is a long-standing problem for a general 
memory friction. Since the effective friction is sufficiently strong (the memory friction
integral diverges) one can tentatively use
a prominent non-Markovian rate theory (NMRT)
result \cite{GroteHynes,HanggiMojtabai,Pollak,HTB90} which is a
generalization of  the celebrated Kramers rate expression \cite{Kramers}. It assumes asymptotically 
an exponential kinetics for the survival probability in one well, 
$P(t)\sim \exp[-R(\mu) t]$,
with the non-Markovian rate
\begin{eqnarray}\label{rate}
R(\mu)=\kappa(\mu)\frac{\omega_0}{2\pi} \exp(-\beta V_0).
\end{eqnarray}
In Eq. \eqref{rate}, $\omega_0=\sqrt{V^{''}(x_{\rm min})/m}=\sqrt{8V_0/(m x_0^2)}$ 
is the bottom attempt frequency, $\exp(-\beta V_0)$ is the Arrhenius factor, $\beta=1/(k_BT)$ is the
inverse temperature, and
\begin{eqnarray}
\kappa(\mu)=\frac{\mu}{\omega_b} \leq 1
\end{eqnarray}
is the transmission coefficient. It invokes the effective barrier 
frequency $\mu$ given by 
the positive solution of equation
\begin{eqnarray}\label{mu}
\mu^2+\mu\tilde\eta(\mu)/m=\omega_b^2,
\end{eqnarray}
where
$\tilde\eta(s)=\int_0^{\infty}\exp(-st)\eta(t)dt$ is 
the Laplace-transformed memory kernel, and 
$\omega_b=\sqrt{|V^{''}(0)|/m}=\omega_0/\sqrt{2}$ is 
the (imaginary) barrier top frequency in the absence of friction. 
We focus below on the case of sufficiently high
barriers, where the Arrhenius factor is small,  $\exp(-\beta V_0)\ll 1$.
Clearly, a good single-exponential kinetics with exponentially 
distributed residence times,
$\psi(\tau)=-d P(\tau)/d\tau$, 
can only be valid for such potential barriers, even 
in the strictly Markovian case.

However, how high is high? Could asymptotically exponential kinetics be attained
for the viscoelastic model considered at all? 
Very important is that  the relaxation within the potential well
is ultraslow and this fact seems to invalidate the non-Markovian rate 
description generally \cite{PRL07}. 
To understand this, let us neglect formally for a while
the inertia effects,  $m\to 0$. Then, the strict power law 
kernel corresponds (in parabolic approximation)
to the relaxation law $\langle \delta x(t)\rangle=
\delta x(0)E_{\alpha}[-(t/\tau_r)^\alpha]$ with the anomalous 
relaxation constant
$\tau_{r}=[\eta_{\alpha}x_0^2/(8V_0)]^{1/\alpha}$, where  
$E_{\alpha}(x)=\sum_{n=0}^{\infty}x^n/\Gamma(1+\alpha n)$
is the Mittag-Leffler function \cite{Min1,PRL07}.
Asymptotically, $\langle \delta x(t)\rangle \propto (t/\tau_r)^{-\alpha}$,
being initially a stretched exponential. 
Precisely  the same relaxation 
law holds also for the CTRW subdiffusion in the parabolic well 
\cite{MetzlerPRL} 
which (along with other similarities for the potential-free case) gave
grounds to believe that these two subdiffusion scenarios are somehow 
related, or similar. 
From the fact of ultraslow relaxation, 
it is quite clear that there cannot be a rate description
even for appreciably high potential barriers, until the relaxation time 
within a potential
well becomes negligible as compared with a characteristic time of escape.
It worth to mention here that the 
non-Markovian rate theory
approach yields a finite rate always, even for the strict power-law kernel 
\cite{Chaudhury}, 
$R_0=R(\omega_r^{(b)})
=\omega_r^{(b)}
\exp(-\beta V_0)/(\sqrt{2}\pi)$, where $\omega_r^{(b)}=
[4V_0/(\eta_{\alpha}x_0^2)]^{1/\alpha}$ is solution of Eq. (\ref{mu})
for $m\to 0$. 

This cannot be, however, always correct.
Indeed, let us introduce {\it ad hoc}
a variable small-frequency cutoff $\nu_c$ such that
$\tilde \eta(s)$ becomes $\tilde \eta(s)=\eta_{\alpha}(s+\nu_c)^{\alpha-1}$.
Then, choosing {\it self-consistently} 
$\nu_c=R(\mu)$ in Eqs. \eqref{rate}-\eqref{mu} (for $m\to 0$) 
one can show that the corresponding 
$\mu$ becomes modified as 
$\mu\to\mu^*=\omega_r^{(b)}[1+\exp(-\beta V_0)/(\sqrt{2}\pi)]^{1/\alpha-1}$.
From this we conclude that the non-Markovian rate expression 
$R(\mu)$ is practically not affected by such a cutoff 
when $\exp(-\beta V_0)\ll 1$. 
This does not mean, however, that all the slowly fluctuating noise contributions
with $\nu_i<R_0$ can be simply neglected. They lead, in fact, to the 
{\it fluctuating rate} description invalidating thereby the 
non-Markovian rate picture.

\subsection{Fluctuating rates: simplest approximation}

The idea is to divide all the noise components 
$\zeta_i$ into the two groups, $\xi(t)=\xi_f(t)+\xi_s(t)$: the fast noise $\xi_f(t)=\sum_{i\leq i_c}\zeta_i(t)$,
which contributes
to the ``frozen'' non-Markovian rate $R(\mu)$, and the slow modes 
which constitute the slowly
fluctuating random force 
$\xi_s(t)=\sum_{i>i_c}\zeta_i(t)$. 
The separation frequency $\nu_{i_c}$ is chosen such that
$\nu_i\leq \nu_{i_c}<R(\mu)$. It depends on the ratio of barrier height and
temperature, as well as $\alpha$.
Furthermore, let's assume
that for the slow $u_i$-modes in 
Eq. \eqref{eq6} one can approximately replace $v(t)$ by its average
zero-value. This is a reasonable approximation because of 
 the dynamics of $v(t)$ is fast on that time scale. 
Then, the corresponding equations for $u_i(t)$ decouple from the particle
dynamics, $u_i\to \zeta_i$, and the corresponding 
stochastic modes can be considered just as an 
external random force. The fast noise agitates the particle trapped (otherwise) 
in the potential
wells leading to the escape events. To a first approximation,
one can regard the slow noise $\xi_s(t)$ be quasi-frozen on the time
scale of such escape events. Then, for high barriers $\beta V_0\gg 1$  
one can use a two-state 
approximation for the overall kinetics with the non-Markovian 
rate $R(\mu^*,\xi_s)$ 
slowly driven in time by $\xi_s(t)$. This slow stochastic force is
in fact  also power-law
correlated. Thus, we are dealing with a typical problem of non-Markovian
dynamical disorder \cite{Goychuk05}. 
Some insight can be obtained 
by using the quasi-static disorder approximation 
\cite{Austin,Zwanzig2,Dewey,Goychuk05} for the averaged kinetics,
\begin{eqnarray}
P_{1,2}(t)=\int_{-\infty}^{\infty}w(\xi_s)\exp[-R_{1,2}(\mu^*,\xi_s)t],
\end{eqnarray}
where $R_{1,2}(\mu^*,\xi_s)$ are the non-Markovian rates for a quasi-static
biasing force $\xi_s$ distributed with the Gaussian probability density
$w(\xi_s)=1/(\sqrt{2\pi}\sigma_s)\exp(-\xi_s^2/(2\sigma_s^2))$
and variance  $\sigma_s^2=k_BT\eta_s$,
where $\eta_s=\sum_{i>i_c}\eta_i$. The calculation
of $\sigma_s$ for $\alpha=0.5$ (or larger) 
shows that
the bias fluctuations are sufficiently small for $\beta V_0\geq 2$, so
that the approximation $R_{1,2}(\mu^*,\xi_s)\approx R(\mu^*)
\exp[\pm \xi_s x_0/(k_BT)]$ can be used. Here, we just assume that
the rms of potential barrier modulations $\pm \sigma_s x_0$ is small
against $V_0$. Since the influence of slow modes on the effective
barrier frequency $\mu$ is exponentially small for high barriers (see above),
one can replace $R(\mu^*)$ with $R(\mu)$. This finally yields
\begin{eqnarray}\label{theory}
P_{1,2}(t)\approx \int_{0}^{\infty}\exp[-R(\mu)t \;y]W(y)dy,
\end{eqnarray}
where $W(y)=1/(\sqrt{2\pi}d y)\exp[-\ln^2(y)/(2d^2)]$ 
is the probability density of log-normal distribution with width 
$d=\sqrt{\eta_s x_0^2/(k_B T)}$. 
The corresponding mean residence time (MRT) and the relative standard deviation,  
$\delta \sigma = \sqrt{\langle \tau^2\rangle - 
\langle \tau\rangle^2}/\langle \tau\rangle \nonumber$, are:  
\begin{eqnarray}\label{mean}
\langle \tau\rangle =R^{-1}(\mu)\exp(d^2/2),
\end{eqnarray}
 \begin{eqnarray}\label{sigma}
\delta \sigma= 
\sqrt{2\exp(d^2)-1} =\sqrt{2[\langle \tau\rangle R(\mu)]^2-1}.
\end{eqnarray}
To characterize non-Markovian kinetics, one can introduce also a time-dependent 
rate $k(t)=-d\ln P(t)/dt$ which decays asymptotically to zero for any finite width $d$
within this approximation.

\begin{figure}
\vspace{0.2cm}
\centerline{\includegraphics[width=7.5cm]{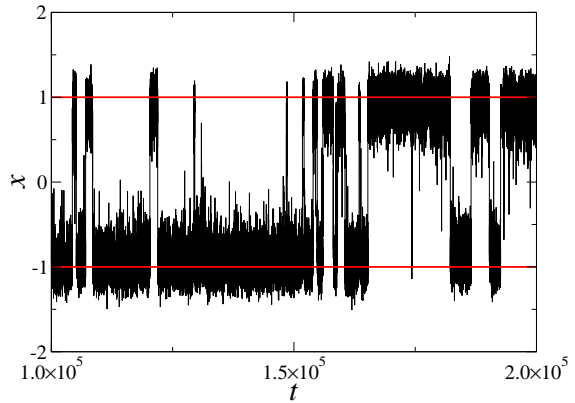}}
\caption{
(Color online) A sample trajectory of bistable transitions for $\beta V_0=6$. 
Time is given in 
units of $\tau_r^{(b)}=[\eta_{\alpha}x_0^2/(4V_0)]^{1/\alpha}$ 
and coordinate in units of $x_0$. The residence
time distributions are extracted by using the two thresholds (red lines).
Stochastic Heun algorithm with the time step $\Delta t=10^{-3}$ and
Mersenne Twister pseudo-random number generator, combined with
the Box-Muller algorithm are used. Markovian embedding with 
$N=16$ is used as described in the text. 
$\alpha=0.5$, $r=10$, $b=10$, $\nu_0=1000$.
}\label{Fig3}
\end{figure}

The resulting physical picture becomes clear:
Fast Ornstein-Uhlenbeck components with $\nu_i>R(\mu)$ 
participate in forming the non-Markovian rate $R(\mu)$, 
while the slow ones lead to a stochastic modulation of this rate in time. 
This implies the following main features which are confirmed further by
a numerical study: 
(i) Both the mean residence time in a potential well, and all the higher
moments exist. 
(ii) Anti-correlations between the alternating
residence time intervals in the potential wells emerge along with a profoundly bursting 
character of the trajectory
recordings, cf. Fig.~\ref{Fig3}.  Indeed, during the time of a quasi-frozen stochastic tilt 
many transitions occur between the potential wells. 
The shorter time in the (temporally) upper well
is followed by a longer time in the lower well. This yields 
anti-correlations (cf. Fig.\ref{Fig4}). Moreover, 
many short-living transitions into the temporally upper
well occur, which appears as bursting (cf. Fig. \ref{Fig3}). 
The subsequent sojourns in one potential well are also positively
correlated for many transitions (not shown). 
(iii) The escape
kinetics is clearly non-exponential, see in Fig. \ref{Fig5} and below.
(iv) The corresponding power spectrum of bistable fluctuations has a
complex structure with several different $1/f$ noise low-frequency 
domains (cf. Fig. \ref{Fig6}); 
(v) The higher is potential barrier, the smaller
is the rms of slow rate fluctuations. The last circumstance implies that
for very high barriers the exponential escape kinetics with non-Markovian rate
in (\ref{rate})-(\ref{mu}) will be restored, cf. Fig. \ref{Fig10}.
For small $\alpha$, this however can require very high barriers and
be practically unreachable.

\subsection{Numerical results}

Let us compare now these theoretical predictions with numerical results.
The time is scaled in this section in the units of 
$\tau_r^{(b)}=1/\omega_r^{(b)}=[\eta_{\alpha}x_0^2/(4 V_0)]^{1/\alpha}$,
which is the anomalous relaxation constant for the inverted parabolic barrier
in the overdamped limit, 
and the role of inertial effects
is characterized by the dimensionless parameter 
$r=\omega_b\tau_r^{(b)}$.  
The used
$r=10$ corresponds to the overdamped limit in the case of normal diffusion. 
For the used $16$-exponential approximation of the memory kernel, it yields
$\mu\approx 0.999$ in Eq. \eqref{mu} which is very close to 
$\mu=1$ corresponding to the
formal overdamped limit, $m\to 0$, with the transmission coefficient 
$\kappa=1/r=0.1$. 
Despite this fact, some inertia effects for the intra-well 
relaxation dynamics are still present. Generally, it is important
to include such effects for a power law memory kernel \cite{Burov}.
We performed
simulations of very long trajectories (from $5\cdot 10^3$ to $10^6$ 
transitions between wells) achieving statistically trustful results in each presented
case. A sample of stochastic 
trajectory for $\beta V_0=6$ is shown in Fig. ~\ref{Fig3}. 
The bursting
character is clear \cite{Goychuk05}, 
indicating also slow tilt fluctuations. 

\begin{figure}[t]
\centering
\includegraphics[width=7.5cm]{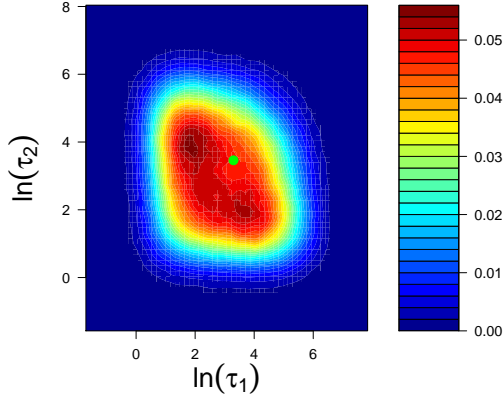}
\caption{(Color online)
Two-dimensional distribution of the log-transformed residence
times in the potential wells for $\beta V_0=2$. The green symbol
corresponds to the non-Markov rate theory 
result (for the studied $16$-exponential expansion of the power-law
memory kernel). The orientation and the structure of the 
plateau of maximal probabilities manifest anti-correlation
of the residence times in two potential wells. This feature is completely
beyond the non-Markovian rate theory. Statistical software R \cite{R} 
is used for data analysis and to produce the plot. \\
 } 
\label{Fig4}
\end{figure}

\begin{figure}[t]
\centering
\includegraphics[width=7.5cm]{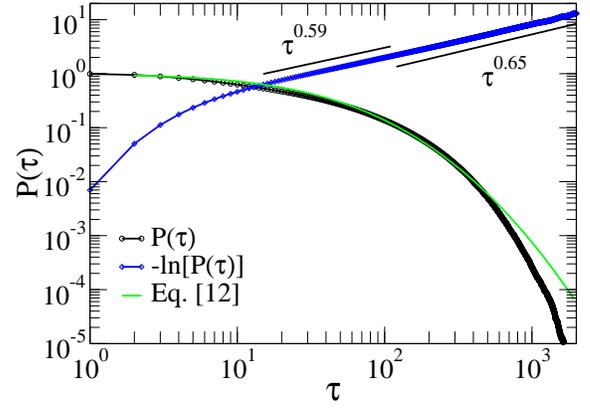}
\caption{(Color online)
Survival probability in one potential well versus time 
(in units of $\tau_r^{(b)}=[\eta_{\alpha}x_0^2/(4V_0)]^{1/\alpha}$)
for $\beta V_0=2$.  The asymptote is stretched-exponential with 
$\gamma\approx0.65$. 
Fit with Eq. \eqref{theory}
is done using $d=0.81$ and $R=2.67\cdot 10^{-2}$. These values were 
derived from the numerical
data using the first two moments of the numerical distribution and 
Eqs. \eqref{mean}, \eqref{sigma}. 
} 
\label{Fig5}
\end{figure}

\begin{figure}[t]
\centering
\includegraphics[width=7.5cm]{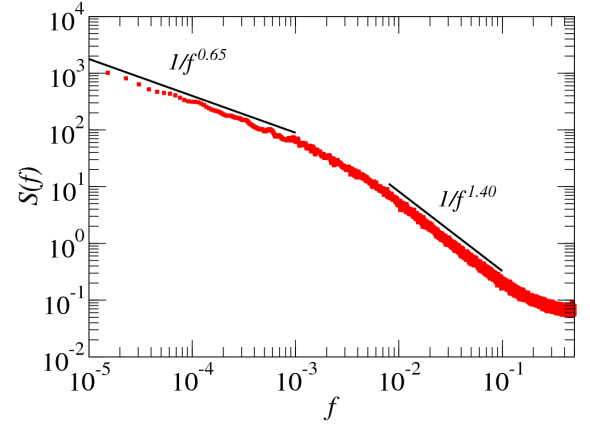}
\caption{(Color online) 
The low-frequency part of the power spectrum of fluctuations for $\beta V_0=2$.
The $1/f^{\gamma}$ feature for smallest frequencies 
is defined chiefly by the stretched-exponential
asymptotics of the survival probability, cf. Fig.~\ref{Fig3}. 
The correlogram method along with the 
SSA-MTM Toolkit for Spectral Analysis 
\cite{Ghil} is used to produce the spectrum.  
\\ \\} 
\label{Fig6}
\end{figure}

To extract the residence
time distributions (RTDs) in the wells, $\psi_1(\tau_1)$ and 
$\psi_2(\tau_2)$, and their joint distribution $\psi(\tau_1,\tau_2)$,
two thresholds were set at the minima of the potential wells,
cf. Fig. \ref{Fig3}. Fig.~\ref{Fig4} displays the the joint distribution 
$\psi(\ln(\tau_1),\ln(\tau_2))$ of the logarithmically transformed  
residence times for $\beta V_0=2$. Two facts  are self-evident: (1) the transformed 
distribution is not sharply
peaked and spreads over several time decades; (2) the subsequent residence times
in two potential wells are significantly anti-correlated. The normalized covariance 
between $\tau_1$ and $\tau_2$ is $c(\tau_1,\tau_2)\approx -0.116$, and
between the logarithmically transformed variables $c(\ln \tau_1,\ln\tau_2)
\approx -0.19$. The mean residence time is approximately 
$\langle\tau_{1,2}\rangle\approx 52$ with the relative standard deviation 
$\delta \sigma_{1,2}=\sqrt{\langle \tau_{1,2}^2\rangle - 
\langle \tau_{1,2}\rangle^2}/\langle \tau_{1,2}\rangle\approx 1.69$, whereas
the non-Markovian rate theory yields
$\tau_{\rm NM}=1/R(\mu)\approx
32.9$. This value essentially underestimates MRT, but it lies not far away from the {\it extended}
region of most probable $\ln\tau_{1,2}$(see the green symbol in Fig. ~\ref{Fig4}).
Furthermore, the distribution of the residence times in each potential well, 
$\psi(\tau)=-\dot P(\tau)$, is profoundly
non-exponential, with a complex kinetics being mostly stretched-exponential,
$P(\tau)\sim\exp[-(\tau/\tau_{\gamma})^\gamma]$. The power $\gamma$ 
slightly varies in time and reaches asymptotically  
$\gamma\approx 0.65$, as indicated by a straight line trend 
for $-\ln[P(\tau)]$ on the doubly-logarithmic plot in Fig. \ref{Fig5}. 
Generally, the asymptotic value of 
 $\gamma$ is bounded as
$\alpha\leq \gamma<1$ and depends on 
the ratio $V_0/k_BT$. 
The formally defined time-dependent non-Markovian 
rate decays  
to zero as $k(t) \propto 1/t^{1-\gamma}$ and 
the corresponding power spectrum of fluctuations in Fig. ~\ref{Fig6} 
displays 
a complex $1/f^{\gamma}$-noise pattern  with the same
$\gamma\approx 0.65$ at lowest frequencies. 
Overally, the non-Markovian rate theory approach is clearly not applicable for such a barrier.

\begin{figure}[t]
\centering
\includegraphics[width=7.5cm]{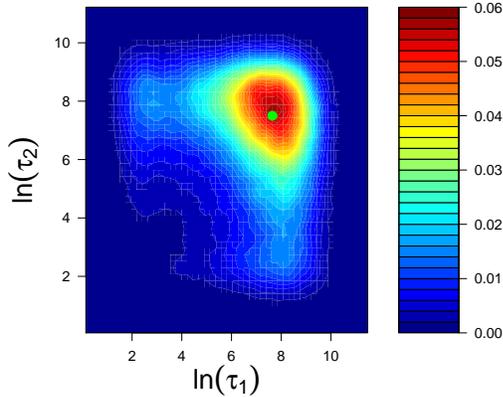}
\caption{(Color online)
Two-dimensional distribution of the log-transformed residence
times in the potential wells for $\beta V_0=6$. The non-Markovian
rate theory result (green symbol) yields the most 
probable value of $\ln(\tau_{1,2})$. Kinetics is, however,
clearly non-exponential, even asymptotically.
 } 
\label{Fig7}
\end{figure}  

  
\begin{figure}[t]
\vspace{0.5cm}
\centering
\includegraphics[width=7.5cm]{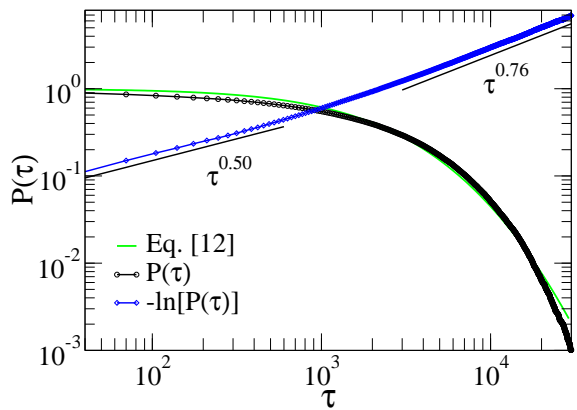}

\caption{(Color online) 
Survival probability in one potential well versus time (in units of $\tau_r^{(b)}$)
for $\beta V_0=6$. Asymptotics is stretched exponential with 
$\gamma\approx0.76$. Initial kinetics is stretched
exponential with $\gamma\approx\alpha$. Fit with Eq. \eqref{theory}
is done using $d=0.634$ and $R=4.48\cdot 10^{-4}$. These values were 
derived from the numerical
data using the first two moments of the distribution and 
Eqs. \eqref{mean}, \eqref{sigma}. 
} 
\label{Fig8}
\end{figure}  

\begin{figure}[t]
\vspace{0.5cm}
\centering
\includegraphics[width=7.5cm]{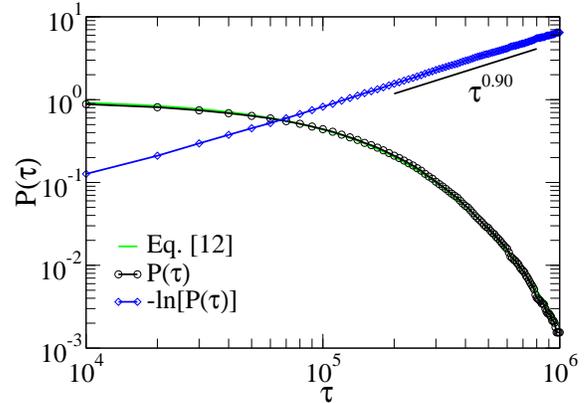}
\caption{(Color online) 
Survival probability in one potential well versus time (in units of $\tau_r^{(b)}$)
for $\beta V_0=10$. Using a fit to the survival
probability, the whole distribution is well described by a stretched
exponential with  $\gamma\approx 0.90$. The maximum likelihood fit to
the numerical data 
yields but a slightly different value $\gamma\approx 0.85$, cf. in Fig. 
\ref{Fig10}, indicating that the whole distribution is not 
properly stretched exponential. It is rather described
by Eq. \eqref{theory} with $d=0.335$ and $R=8.26\cdot 10^{-6}$.
These values were derived from the numerical
data using the first two moments of the distribution and 
Eqs. \eqref{mean}, \eqref{sigma}. 
} 
\label{Fig9}
\end{figure}

\begin{figure}[t]
\vspace{0.5cm}
\centering
\includegraphics[width=7.5cm]{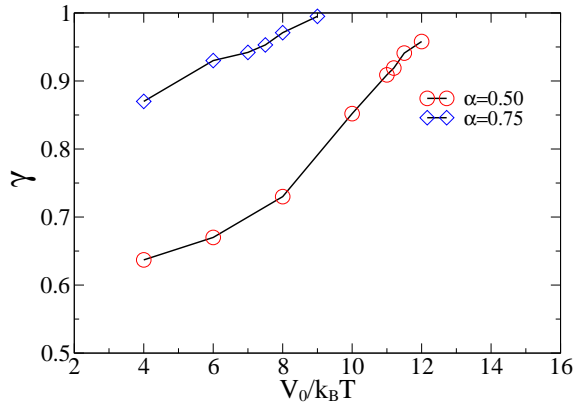}

\caption{(Color online) Power exponent of a (single) stretched exponential
fit to the overall residence time distribution versus the barrier height for
two different subdiffusion exponents $\alpha$. The maximum likelihood
approach is used to derive $\gamma$ from data. Notice that this $\gamma$
does not correspond to the asymptotic $\gamma$ in the previous 
figures and text, 
but rather
presents an average of $\gamma(t)$ changing slowly in time.
This subtle difference diminishes when $\gamma$
approaches one. For $\alpha=0.75$, $C_{0.75}=1.885$; other parameters
in the kernel approximation are the same as for $\alpha=0.5$.  
For $V_0$ exceeding some borderline value $V_c(\alpha,T)$, 
$\gamma$ tends to one
and kinetics becomes gradually single exponential.
} 
\label{Fig10}
\end{figure}

The qualitatively similar features remain also for some higher potential 
barriers (or lower temperatures), e.g. for $\beta V_0=6$.
In this case, numerically  $\langle \tau_{1,2}\rangle \approx 2710$, and
$\delta \sigma_{1,2}\approx 1.41$, whereas
the non-Markovian rate theory yields  
$\tau_{\rm NM}\approx 1794$ with $\ln(\tau_{\rm NM})\approx 7.49$. This 
NMRT result compares, however, now well against the most probable $\ln(\tau_{1,2})$
in Fig. \ref{Fig7}. This provides one of important results: 
Even if the non-Markovian rate theory is still not applicable, it can predict
remarkably well the most probable logarithm of residence times. 
The kinetics remains asymptotically stretched-exponential even for such a high barrier 
with $\gamma$ increased to $\gamma\approx 0.76$ (cf. Fig. \ref{Fig8}).
However, the region of
most probable $\ln \tau_{1,2}$ shrinks further with increasing 
$\beta V_0$ and the non-Markovian rate theory describes ever better both the most
probable $\ln \tau_{1,2}$, and the (logarithm of)  mean residence time which start
to merge as it should be for a single-exponential RTD. Already for 
$\beta V_0=10$ 
the whole distribution is well approximated 
by the  stretched exponential with 
$\gamma\approx 0.90$, Fig. \ref{Fig9}. 
Clearly, for ever higher
barriers 
the transition kinetics becomes gradually single-exponential. 
This happens when the
barrier height exceeds some characteristic value $V_c(\alpha,T)$ which depends on $\alpha$
and temperature,
cf. Fig. \ref{Fig10}. Since there is no a precise threshold, the
definition of $V_c$ is rather ambiguous. A working criterion for defining $V_c$ can be, e.g., 
that the rate description is achieved within some error bound, e.g. $1\%$
for deviation of $\gamma$ from unity. 
 
For $\beta V\gg \beta V_c$, the overall escape kinetics is well-described
by the non-Markovian rate theory. For $\alpha=0.5$ it is very
difficult to obtain a good statistics of transitions to find precisely
this borderline. For the maximal 
value $\beta V_0=12$ used in our simulations 
the maximum likelihood fit
with a stretched exponential (Weibull) distribution yields 
$\gamma=0.952 \pm 0.034$. This value of 
$\beta V= 12$ provides an estimate for $\beta V_c$
from below for $\alpha=0.5$. It approximately delimits
the borderline between the applicability of 
NMRT for $\alpha=0.5$
and our treatment beyond it. The lower is $\alpha$, the higher is 
borderline $\beta V_c(\alpha)$, and vice versa. 
For example, for $\alpha=0.75$, $\beta V_c$ reduces to
about $\beta V_c \approx 9$. For these parameters, the maximum likelihood
fit of the numerical data with the single exponential
distribution yields the rate $R=(2.864 \pm 0.065)\cdot 10^{-5}$
which only slightly deviates (about 3\%  of error) 
from the corresponding non-Markovian rate theory result 
$R(\mu)\approx 2.775\cdot 10^{-5}$. And for $\beta V_0=10$, the
maximum likelihood fit yields $R=(1.056\pm 0.034)\cdot 10^{-5}$ ($\alpha=0.75$)
which almost agree within statistical errors with the non-Markovian
rate theory result $R(\mu)=1.021\cdot 10^{-5}$.
This provides a spectacular
confirmation of both the non-Markovian rate theory for very high potential
barriers, and the reliability of our numerics, as well as the
physical picture of anomalous escape developed in this work. 
On the contrary, for $\alpha<0.5$, e.g. 
$\alpha =0.25$, $\beta V_c$ can be so high that the non-Markovian 
rate theory limit will never be reached for realistic barrier heights. 
 Both our theoretical argumentation
and the numerical results show that this borderline is fuzzy and
and the rate description is restored gradually. The tendency in 
Fig. \ref{Fig10} is, however, obvious. 

The quasi-static disorder approximation cannot  describe quantitatively
the numerical results  for a broad range of parameters (rate disorder is yet dynamical, in spite of a quasi-infinite
autocorrelation time \cite{Goychuk05}). 
Nevertheless, it captures the essential physics (cf. Figs. \ref{Fig5},\ref{Fig8})
and becomes ever better with increasing the barrier height, cf. Fig. \ref{Fig9}.
The agreement in this figure proves that our theory is essentially correct
predicting the correct trend with increasing  the barrier height, at least 
for $\alpha>0.5$. Indeed, with increasing the barrier height, or lowering
the temperature the averaged escape time  increases exponentially with 
$\beta V_0$ and, therefore, ever more slow noise components $\zeta_i$ 
contributes to the non-Markovian rate and ever less such components contributes
to  fluctuation of  this rate. For this reason, the 
root mean-squared amplitude
of the slow (in our terminology) 
stochastic force $\xi_s(t)$ gradually diminishes. For some 
characteristic $\beta V_c$, which clearly depends on $\alpha$,
it becomes negligible and the single-exponential kinetics is then 
approximately restored. The corresponding rate is given by the 
non-Markovian rate theory. 

The physical picture developed in this work is very different from the
previous attempts  in Refs.
\cite{Chaudhury,PRL07} to solve the problem of anomalous 
escape utilizing different approximations.
The Ref.
\cite{Chaudhury} focuses on the subdiffusive transmission through the
parabolic barrier. It predicts that asymptotic rate
$k_{\infty}=-\lim_{t\to\infty}  d \ln P(t)/dt$ {\it always} exists and
is given precisely by the non-Markovian rate theory result in Eqs.
(\ref{rate})-(\ref{mu}).  This is clearly not correct for
$V<V_c(\alpha,T)$. Strictly speaking,  $k_{\infty}=0$ always, even
for $V>V_c(\alpha,T)$ for a strictly power law memory kernel.
However, for $V\gg V_c(\alpha,T)$, the shape factor of 
Weibull distribution  
$\gamma$  equals approximately one, $\gamma\approx 1$,  
and the rate description provides a
good approximation. The higher is $V_0$, the better is this
approximation, cf. Fig. \ref{Fig10}. The Ref.
\cite{PRL07} focuses on the escape of a massless particle ($m\to 0$)
from a parabolic potential well with a sharp cusp-like cutoff, utilizing the
non-Markovian Fokker-Planck equation (NMFPE) of  Refs.
\cite{Adelman,GroteHynes,HanggiMojtabai}. This NMFPE is exact for the
parabolic potential, being but only approximate for a parabolic
potential with cutoff. The better is the Gaussian approximation, the
better should be the corresponding description, which implies high
potential barriers  $\beta V_0\gg 1$. The theory in \cite{PRL07} cannot be
compared directly with the present one (different potentials, zero
mass particle in  \cite{PRL07}, expansion of the power law kernel into
a {\it finite}  sum of exponentials here), and the {\it extrapolation} of
some main results in Ref. \cite{PRL07} on a more realistic case here, 
would lead to the conclusions which are at odds with
the present theory. In particular, the  theory in \cite{PRL07}
predicts (for a strict power law kernel, without inertial effects)  
that the escape kinetics is
asymptotically a power law, being only initially stretched
exponential, and that the corresponding effective power law  exponent
tends exponentially to zero with increasing $\beta V_0$. This means
that the particle  becomes strongly localized with increasing  the
barrier height, and the corresponding kinetics becomes ever more
abnormal. On the contrary, the present theory predicts that the escape
kinetics tends to a normal one, even if it decelerates dramatically. 
For a memory kernel with cutoff, the
theory in Ref.  \cite{PRL07} predicts that with increasing the
barrier  height the kinetics does become normal, when the memory
cutoff becomes shorter than the mean escape time. This prediction 
concords with the present theory. The difference is however that
the physical picture developed in this work suggests that the escape
kinetics can be also approximately exponential when the memory cutoff
largely exceeds the mean escape time.  
To conclude, the  theory
in this work is more physical.  It overcomes the previous attempts to
solve the very nontrivial problem of subdiffusive escape  by taking a quite different
road of multidimensional Markovian embedding 
and  it is confirmed by numerics. 

The fact that the escape kinetics tends to a single-exponential
with increasing the barrier height 
does not
mean, however, that the diffusion becomes normal in the periodic
potentials, as one might naively think in analogy with the CTRW theory. 
As a matter of fact, asymptotically such a diffusion cannot be faster that one in the
absence of potential, i.e. $\langle \delta x^2(t)
\rangle\propto t^{\alpha}$. Therefore, we expect here new surprises.

\section{Subdiffusion in periodic potentials}

\begin{figure}[t]
\centering
\includegraphics[width=7.5cm]{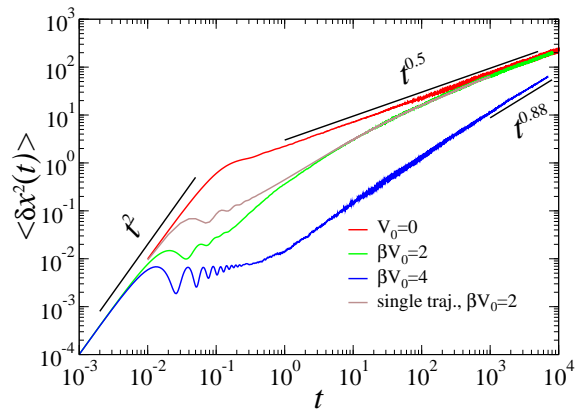}
\caption{(Color online) Diffusion in washboard potentials.
The distance is measured in units of period $L$ and the time in units of 
$\tau_{\beta}=(\beta L^2\eta_{\alpha})^{1/\alpha}$. $\alpha=0.5$ and the 
damping parameter $r_{\beta}=\tau_{\beta}/\tau_{\rm bal}=10$
with $\tau_{\rm bal}=L/v_T$ and $v_T=\sqrt{k_B T/m}$. 
Particles are initially localized at $x=0$ with Maxwellian
distributed velocities. The number of particles $n=10^4$ is used in simulations
(stochastic Heun algorithm, time step $\Delta t=10^{-4}$) for 
the ensemble averaging. 
Initially, the diffusional broadening is always ballistic due to inertia effects. 
In potentials, this regime is followed
by transient rattling oscillations due to a combination of the cage effect and the
influence of the potential force. On the space scale $\delta x>L$, 
diffusion becomes ultraslow 
$\langle \delta x^2(t)\rangle \propto t^{\alpha_{\rm eff}(t)}$ with a slowly changing
$\alpha_{\rm eff}(t)$ which depends on the potential height. 
Asymptotically, $\alpha_{eff}(t\to\infty)\to 0.5$. 
This universal asymptotics is almost reached for $\beta V_0=2$. 
} 
\label{Fig11}
\end{figure}

We consider a common type washboard potential $V(x)=-V_0\cos(2\pi x/L)$
with the spatial period $L$.
To study the influence of periodic potential on free subdiffusion, it is convenient
to scale now the time  in the units of 
$\tau_{\beta}=(\beta L^2\eta_{\alpha})^{1/\alpha}$, as in Fig. \ref{Fig2},
which does not depend on the barrier height $2V_0$. It takes time about 
$\tau_{\beta}$ to subdiffuse freely over the distance about $L$. 
Indeed, the numerical simulations
 for $\beta V_0=2$
delivers a surprise indicating, see in Fig. ~\ref{Fig11}, 
that the presence of periodic potential does 
not influence subdiffusion
asymptotically. This seems to agree with a theory 
in Refs. \cite{Chen,WeissBook} which, however, cannot be invoked directly because of 
it relates to a fully quantum case, where the tunneling effects generally contribute. From this
agreement we can, however, conclude that this surprising effect is certainly {\it not}
of the quantum nature in the quantum case, but reflects the {\it anti-persistent} character of our
viscoelastic subdiffusion which is purely classical. Namely, 
it is not diverging MRT, but extremally long-lived displacement (and velocity) 
anti-correlations which are responsible for the observed anomalous diffusion 
behavior in viscoelastic media. It must be emphasized, however, this asymptotical 
regime is achieved through 
very long transients with a time-dependent
$\alpha_{\rm eff}(t)$ gradually approaching $\alpha$, cf. 
Fig. ~\ref{Fig11}. This feature is  totally beyond any asymptotical analysis like one 
in Ref. \cite{Chen}. 
The potential barrier height does generally matter and it
{\it strongly} influences the whole time-course of diffusion.
After a short ballistic 
stage 
followed by decaying coherent oscillations due to a combination of 
inertial and cage effects \cite{Burov}
in a potential, 
the diffusion can look initially close to normal (as for $\beta V_0=4$ in 
Fig. ~\ref{Fig11}).
This is due to a finite mean residence time in a potential well.
However, it slows down and turns over into subdiffusion. 
The borderline of free subdiffusion
cannot be crossed, cf. Fig. \ref{Fig11}.

Very important is that the free viscoelastic subdiffusion is ergodic,
in agreement with \cite{Deng}. The results
of the ensemble averaging with $n=10^4$ particles coincide with the
time averaging for a single particle done in accordance with Eq. (\ref{A5}),
see in Fig.~\ref{Fig2}(a). However, 
in the periodic potential a strong deviation 
from the ergodic behavior takes place on the averaged time scale of the escape to 
the first neighboring potential wells, cf. Fig. \ref{Fig11}. This reflects
anomalous escape kinetics as discussed above. Nevertheless, on a larger time-scale  
the single-trajectory averaging 
and the ensemble averaging yield again the same results. 
All this is in  striking contrast 
with the CTRW-based subdiffusion, both free 
\cite{Hughes,Shlesinger,He,Lubelski} 
and in periodic potentials \cite{Rapid06,Heinsalu06}. 
In this respect,
the benefit which subdiffusional search  can bring for functioning 
the biological
cells machinery \cite{Golding,Guidas} will not be questioned by the 
weak ergodicity breaking, as it might be in the case of 
CTRW-based subdiffusion. Indeed, the weak ergodicity breaking is related 
to a spontaneous localization of CTRW-subdiffusing particles -- i.e., 
a  portion
of them does not move at all (individual diffusion coefficient is close to zero), 
while other diffuse  with an inhomogeneously distributed
normal diffusion coefficient \cite{He,Lubelski} which depends on the total observation
time ${\cal T}$,
 even if the particles are totally identical.
Numerically, or in a real experimental setup the ergodic behavior
is achieved in the case of viscoelastic subdiffusion 
for very long ${\cal T}$ only, see also in \cite{Deng}. 
So, to check the ergodicity  for
times until $t=10^3$ we run a single trajectory for overall ${\cal T}=10^7$ 
(this corresponds roughly to sampling over $n=10^4$ copies,
in analogy with the ensemble averaging).
For a much smaller ${\cal T}$, the difference between the time and ensemble averagings
becomes sizeable. Hence, experimentally one can yet observe 
broadly distributed subdiffusion coefficients, especially if both the particles
and their environments are subjected to statistical variations \cite{Golding}.
However, differently from the CTRW subdiffusion, a particle will never 
get spontaneously
trapped for the time of observation.  This provides a true benchmark to distinguish
between these two very different subdiffusion mechanisms experimentally.

\section{Summary}

Our main results are of profound importance for the anomalous diffusion and rate theory settling a
long-standing and controversial issue with conflicting results 
of different approaches, and different approximations. 
In particular, we prove that subdiffusion does not require
principally a divergent mean residence time in a finite spatial domain, which makes it less
anomalous when the anti-persistent, viscoelastic mechanism is at work. 
Moreover, we substantiate  the validity of the celebrated
non-Markovian rate theory result \eqref{rate}-\eqref{mu} 
for very high potential barriers ($\beta
V_0> 12$ for $\alpha=0.5$ and $\beta
V_0>\beta V_c\approx 9$ for $\alpha=0.75$ ) even for a strict power law memory kernel, where it
was not expected to work because of an ultraslow relaxation 
within a potential well. However, for small $\alpha<0.5$ the corresponding
borderline value $\beta V_c(\alpha)$ can be so high that this regime becomes practically 
unreachable, at least for numerical simulations.
Surprisingly, the non-Markovian theory result remains useful also for intermediate barriers, 
$2 k_B T< V_0< V_c(\alpha,T)$, where it predicts
the most probable logarithm of dwelling times.
Here, the physics is well described by slowly fluctuating
non-Markovian rates.
For small barriers, $\beta V_0<1$, and for other models, e.g. 
when the bottom of potential well becomes more extended and flat, like 
the potential box in \cite{PNAS02}, the fluctuating rate approach also  
loses its heuristic power. Then, the sluggish approach from the bottom 
to the barrier crossing region determines the transition kinetics. 
Even in the case of normal diffusion,  
different power-law kinetic regimes emerge \cite{PNAS02} 
and the anomalous intrawell diffusion 
can change the corresponding power-law exponents, 
as modeled within the CTRW approach in Ref.
\cite{Goychuk04}. 

One of generic results is that the CTRW subdiffusion and the 
GLE subdiffusion are profoundly different, in spite of some
superficial similarities. 
Subdiffusion in periodic potentials highlights the 
differences especially clear. Surprising is the finding that asymptotically
the GLE subdiffusion is not sensitive to the barrier height, even if imposing
a periodic potential does strongly affect the overall time-course of diffusion, and for
a high potential barrier subdiffusion can look normal 
on a pretty long time interval. However, it slows down and asymptotically
approaches the borderline of free subdiffusion. 
 Such subdiffusion operates within
a quite different (as compared with CTRW) physical mechanism based on
the anti-persistent long-range correlations 
and not on the residence time distributions with  divergent
mean. 

We believe that our results require to look anew on the 
theoretical interpretation of 
experimental subdiffusion
results in biological applications, where the issue of ergodicity
can be crucial. They provide some additional theoretical support  for 
the viscoelastic subdiffusion mechanism. A further detailed study
is, however, necessary. 
To conclude, our work consolidates 
viscoelastic subdiffusion and fractional Brownian 
motion with the non-Markovian rate theory and fluctuating rate (dynamical
disorder) approaches. It also agrees with  
the already textbook view [see, e.g. in \cite{Nelson} (pp. 380-382)], 
of the unusual kinetics as one with 
quasi-frozen and quasi-continuous conformational 
substates, as it was pioneered in biophysical applications by 
Austin {\it et al.}\cite{Austin}.

\begin{acknowledgments}
Support of this work by the DFG-SFB-486 and by Volkswagen-Foundation (Germany),
as well as useful discussions with P. Talkner and P. H\"anggi are gratefully 
acknowledged.
\end{acknowledgments}

\appendix
\section{Exact solution of the potential-free problem 
versus Monte Carlo simulations}

In this Appendix, we discuss numerical errors by comparison of 
the approximate results with 
the exact solution of subdiffusion problem in the absence of
any potential. This exact solution is well-known \cite{Wang,Kupferman}.
Assuming initial velocities to be thermally distributed,
it reads 
\begin{eqnarray}\label{exact}
\langle \delta x^2(t)\rangle  =  2v_T^2\int_0^t H(t')dt', 
\end{eqnarray}
where
\begin{equation}
H(t)=\int_0^{t} K_v(\tau) d\tau  
\end{equation}
is the integral of 
normalized equilibrium velocity autocorrelation function 
$K_v(\tau):=\langle v(t+\tau)v(t)\rangle/\langle v^2(0)\rangle$ with
$\langle v^2(0)\rangle=v_T^2=k_BT/m$. It has the Laplace-transform
\begin{equation}\label{Kv}
\tilde K_v(s) = \frac{1}{s+\tilde \eta(s)/m}\;.
\end{equation}
Accordingly, the Laplace-transform of the coordinate variance$, \langle \delta \tilde{ x^2}(s)\rangle:=
\int_0^{\infty}\exp(-s t)\langle \delta x^2(t)\rangle dt$,
is
\begin{eqnarray} \label{free}
\langle \delta \tilde x^2(s)\rangle =
\frac{2 v_T^2}{s^2[s +\tilde \eta(s)/m]}\;,
\end{eqnarray}
For the strict power-law
kernel,  Eq. (\ref{free}) becomes
$\langle \delta \tilde x^2(s)=2/[s^3/r^2+s^{1+\alpha}]$
with the distance measured  in some arbitrary units $L$ and time in the units of 
$\tau_{\beta}=(L^2\eta_{\alpha}/k_BT)^{1/\alpha}$.
This result can be inverted to the time domain in terms of the generalized 
Mittag-Leffler functions, see e.g. in \cite{Burov}. 
However, both for the exact memory kernel
and for its approximation in Eq. (\ref{approx}),
it is convenient
to invert Eq. (\ref{free}) numerically using the 
Gaver-Stehfest method \cite{Stehfest} 
with arbitrary numerical 
precision, as done e.g. in Ref. \cite{GoychukChemPhys}
for a different problem.
The results thus obtained are 
numerically exact and the algorithm is very fast. 
They are compared
against the results of the Monte-Carlo simulation of 
Eqs. (\ref{embedding})-(\ref{eq6}) in 
Fig. \ref{Fig2}. In these simulations, we used both
the ensemble averaging over $n=10^4$ trajectories and a
 time averaging for single trajectory
defined by 
\begin{eqnarray}\label{A5}
\langle \delta x^2(t)\rangle_{\cal T} =\frac{1}{{\cal T}-t}
\int_0^{{\cal T}-t}[x(t+t')-x(t')]^2dt'\,
\end{eqnarray}
for a very large time window ${\cal T}\gg t$.
The relative error in Fig. \ref{Fig2}(b) is
calculated as $\delta=|\langle \delta x^2_{\rm exact}(t)\rangle
-\langle \delta x^2_{\rm approx}(t)\rangle|/
\langle \delta x^2_{\rm exact}(t)\rangle$, where
$\langle \delta x^2_{\rm approx}(t)\rangle$ is either the
result of numerical solution of stochastic differential
equations (Monte Carlo, with $10^4$ trajectories -- noisy looking
data), 
or  the result of $16$-exponential 
approximation in Eq. (\ref{free}).
The agreement is confirming both for the used
approximation of the memory kernel and for the quality of our stochastic
simulations. The error introduced by the kernel approximation
 is mostly less than
1\%. The well-known phenomenon of logarithmic oscillations \cite{Hughes}
occurs within
this error margin, and, therefore, practically 
does not influence our stochastic numerics, 
which have a typical error of less than  3\% (maximal 5\%). 
Notice that some damped oscillations
in Fig.\ref{Fig11} are of the inertial origin and 
have nothing in common with the logarithmic oscillations seen 
in Fig. \ref{Fig2}(b).
Moreover, for the octave scaling, $b=2$, and for $N=64$ such logarithmic
oscillations are even not present in the variance behavior (not shown).
 The error margin in the variance behavior 
did not become, however, appreciably narrower.
 Therefore, such logarithmical oscillations  practically do not
matter for our numerics and the used $N=16$ embedding 
computationally is even preferred.

It worth to mention that the final time $t=10^4$ in our
simulations with $n=10^4$ trajectories corresponds to about one
week of computational time. Therefore, on our computers the theoretical
limit of free normal diffusion for $t>10^{12}$ for the used $N=16$ embedding 
cannot be reached in principle. The presented data prove that our Markovian 
embedding  is indeed both
of a very good quality, and of practical use.









\end{document}